\documentclass[12pt,a4paper]{article}

\usepackage[a4paper,margin=0.5in]{geometry}
\usepackage[T1]{fontenc}
\usepackage{mathptmx}
\usepackage{amsmath,amssymb,amsfonts}
\usepackage{booktabs}
\usepackage{array}
\usepackage{graphicx}
\usepackage{url}
\usepackage{xcolor}
\usepackage{cite}
\usepackage{multirow}
\usepackage{setspace}
\usepackage{titlesec}
\usepackage{eso-pic}
\usepackage[colorlinks=true,linkcolor=blue,citecolor=blue,urlcolor=blue]{hyperref}
\usepackage{tikz}
\usetikzlibrary{arrows.meta,positioning,shapes.geometric,fit,calc}

\titleformat{\section}
  {\normalfont\bfseries\normalsize}
  {}
  {0pt}
  {\MakeUppercase}
\titlespacing*{\section}{0pt}{2.2ex plus 0.6ex minus 0.2ex}{0.8\baselineskip}

\titleformat{\subsection}
  {\normalfont\bfseries\normalsize}
  {}
  {0pt}
  {}
\titlespacing*{\subsection}{0pt}{1.6ex plus 0.5ex minus 0.2ex}{0.5\baselineskip}

\newcommand{\papertitle}{A unified reconstruction algorithm for reduced-frame structured illumination microscopy}
\newcommand{\paperauthors}{Jingxiang Zhang\textsuperscript{1}, Tianyu Zhao\textsuperscript{1}, Manming Shu\textsuperscript{1}, Keru Mou\textsuperscript{1}, Zheming Zhang\textsuperscript{1}, Yihan Sun\textsuperscript{1},  Mengrui Wang\textsuperscript{1}, Yansheng Liang\textsuperscript{1}, Shaowei Wang\textsuperscript{1}, Ming Lei\textsuperscript{1,*}}
\newcommand{\paperaffiliations}{\textsuperscript{1}MOE Key Laboratory for Nonequilibrium Synthesis and Modulation of Condensed Matter, School of Physics, Xi'an Jiaotong University, Xi'an 710049, People's Republic of China}
\newcommand{\papercorrespondence}{ming.lei@mail.xjtu.edu.cn}
\newcommand{\installpreprintheader}{%
  \AddToShipoutPictureFG{%
    \AtPageUpperLeft{%
      \raisebox{-12pt}[0pt][0pt]{%
        \makebox[\paperwidth][c]{%
          \parbox{0.78\paperwidth}{\centering\fontsize{7}{8}\selectfont
          Preprint manuscript draft; this version compiled \today.\\
          This manuscript has not been peer reviewed.}}}}}}

\newcommand{\makepreprinttitlepage}{%
\thispagestyle{plain}
\vspace*{0.4em}
\noindent{\Large\bfseries Title | \papertitle\par}
\vspace{1.4em}
\noindent\textbf{Authors | } \paperauthors\par
\vspace{1.4em}
\noindent\textbf{Affiliations}\par
\vspace{0.5em}
\noindent\paperaffiliations\par
\vspace{1.0em}
\noindent\textbf{*Correspondence to:} \href{mailto:\papercorrespondence}{\papercorrespondence}\par
\vfill
\clearpage
}

\begin{document}
\installpreprintheader
\makepreprinttitlepage

\section*{ABSTRACT}
Reduced-frame structured illumination microscopy (SIM) is attractive for live-cell imaging because it can improve temporal throughput and reduce photobleaching, but incomplete phase sampling makes reconstruction unstable and computationally demanding. Here we present URA-SIM, a unified reduced-acquisition framework that turns fixed reduced-frame measurements into pipeline- compatible raw stacks through model-consistent phase-domain completion. Instead of solving a large object-level inverse problem or replacing established SIM reconstruction, URA-SIM estimates the missing phase content on the low-dimensional phase-harmonic manifold required by the target modality and then delegates order separation and image formation to classical reconstruction pipeline. This design combines three practical advantages: fidelity from the SIM forward structure, lightweight online computation, and direct
compatibility with existing reconstruction workflows.
For 2D-SIM, URA-SIM uses the first-harmonic phase structure of three-phase SIM to estimate a shared zero-order field and complete missing phase samples by direction-wise harmonic fitting. On calibration and biological 2D-SIM data, reduced-frame reconstructions preserve resolvable structures and remain competitive on COS7 mitochondria comparison data. In live-cell COS7 mitochondria imaging, URA-SIM reconstructs each time point from five acquired raw frames and resolves mitochondrial cristae across different temporal sampling regimes.   Experiments on 3D-SIM and nonlinear SIM further show that the same design principle can be transferred when the phase model and reconstruction-pipeline interface are adapted to the target modality. These results support URA-SIM as a transparent, model-consistent and computationally lightweight route from fixed reduced-frame acquisition to classical SIM reconstruction workflows.

\medskip
\noindent\textbf{Keywords:} structured illumination microscopy, reduced-frame imaging, phase completion

\section{Introduction}
Structured illumination microscopy (SIM) remains one of the most practically useful super-resolution modalities, offering a favorable compromise between resolution enhancement, imaging speed, and phototoxic burden compared with localization-based or depletion-based nanoscopy methods \cite{heintzmann1999laterally,gustafsson2000surpassing,kner2009super,wu2018faster}. This practical appeal, however, comes at a cost: conventional SIM reconstructs each super-resolved frame from multiple phase-shifted raw images acquired under a prescribed illumination protocol — typically nine raw frames for 2D-SIM, fifteen for standard 3D-SIM, and twenty-five or more for nonlinear SIM, where higher-order harmonics must be resolved \cite{gustafsson2008threedimensional, orieux2012bayesian, chen2023simreview}.

This multi-frame requirement is the dominant bottleneck for live-cell and dose-sensitive applications, since temporal resolution and total light exposure scale directly with the number of raw exposures \cite{york2013instant, markwirth2019vigor}. Speed-limit analyses accordingly indicate that further gains must come either from acquiring fewer raw frames or from extracting more information per measurement \cite{strohl2017speedlimits, zhao2025realtimesim}. Importantly, the conventional frame counts above are standard protocols chosen for robust harmonic separation, not immutable physical lower bounds, which leaves room in principle for reduced-frame acquisition \cite{orieux2012bayesian, dong2015resolution, strohl2017speedlimits, lal2018frequency, jin2020dlsim}.

Realizing this potential, however, is non-trivial. Many notable advances have improved SIM along complementary axes — reconstruction fidelity \cite{wen2021hifi}, parameter calibration \cite{qian2023elightpca}, and temporal performance \cite{ma2018structured, xu2024ultra} — but most retain the standard multi-frame protocol and target reconstruction quality rather than the raw-frame burden itself \cite{lukevs2014comparison}. The reduced-frame literature directly addresses this burden, yet it remains intrinsically difficult: complete phase stepping is precisely what keeps classical order separation well conditioned and demodulation reliable. Removing raw frames degrades conditioning, weakens phase diversity, and amplifies sensitivity to parameter mismatch and model error \cite{lal2018frequency,orieux2012bayesian}.

One representative attempt is the Bayesian formulation, which established the feasibility of reduced-frame SIM reconstruction from incomplete phase measurements \cite{orieux2012bayesian, lukevs2014three}. In reduced-frame settings, however, posterior uncertainty and limited phase diversity make reconstruction increasingly sensitive to prior choice and illumination-parameter mismatch, with a direct risk of artifact amplification or reconstruction failure \cite{labouesse2017joint,van20193d,wen2021hifi}. A complementary route is the frequency-domain framework of Lal \emph{et al.}, which treats reduced-image SIM reconstruction in an end-to-end manner \cite{lal2018frequency}. Despite this methodological completeness, reconstruction stability can degrade in practice, and the computational burden is high with limited runtime efficiency, which constrains practical use in live-cell scenarios where reconstruction throughput is critical. 
Deep learning-based SIM reconstruction is also attractive, but it depends strongly on training data and on distributional agreement between training and deployment settings. In reduced-frame SIM, this can lead to limited generalization, reduced interpretability, and a risk of hallucinated or overly smoothed structure, especially when the acquisition protocol or specimen type changes \cite{jin2020dlsim, cheng2022fast, wu2024single}.
These limitations motivate URA-SIM (Unified Reduced-Acquisition SIM), a framework that delivers both high fidelity and high computational performance by using modality-specific completion layers for 2D-SIM, 3D-SIM, and NL-SIM, while remaining aligned with the demands of live-cell SIM imaging.

Here we introduce URA-SIM as a unified reduced-frame reconstruction framework for SIM with three design targets: high fidelity, high computational performance, and practical compatibility with existing reconstruction pipelines. Rather than attempting exact recovery of the missing raw measurements, URA-SIM performs model-consistent phase completion and then delegates final reconstruction to a mature SIM reconstruction pipeline. The central principle is to complete the reduced stack on the phase-harmonic manifold required by the target pipeline, so that reduced acquisition can be integrated without redefining the full reconstruction chain. Specifically, we make four points. First, we define a fidelity-oriented completion model from the SIM forward structure and make its assumptions explicit. Second, we keep the online computation lightweight by combining low-dimensional completion with established reconstruction pipelines, instead of heavy joint inversion at runtime. Third, we present URA-SIM as a unified reduced-acquisition design principle instantiated across 2D-SIM, standard 3D-SIM and NL-SIM by matching each modality's phase-harmonic basis, phase-sampling scheme, and reconstruction-pipeline interface. Fourth, we align URA-SIM with live-cell use by coupling fixed reduced-acquisition protocols to stable, throughput-oriented reconstruction.

\section{Methods}
\subsection{Overview of the URA-SIM workflow}
URA-SIM is a reduced-acquisition SIM reconstruction framework built around phase-domain raw-stack completion. Given a reduced set of phase-shifted raw images, the method estimates the phase-harmonic content required by the target SIM modality, synthesizes the missing phase samples, and forms a completed raw stack with the frame layout expected by a conventional reconstruction pipeline. The pipeline then performs the final SIM reconstruction using its standard calibration, order-separation, and image-formation procedures. In this design, URA-SIM does not replace the SIM reconstruction engine; instead, it acts as a model-consistent interface that converts reduced-frame measurements into a pipeline-compatible raw stack. The completed stack size is determined by the acquisition format of the target modality: nine frames for 3-direction, 3-phase 2D-SIM, fifteen frames for standard 3D-SIM, and twenty-five frames for the nonlinear SIM experiment considered here. The completion model is therefore modality-specific, but the workflow is shared across cases: estimate the relevant phase-harmonic representation from the reduced measurements, complete the missing phase samples, and reconstruct the completed stack with the corresponding SIM pipeline.

\begin{figure}[t]
\centering
\includegraphics[width=\textwidth]{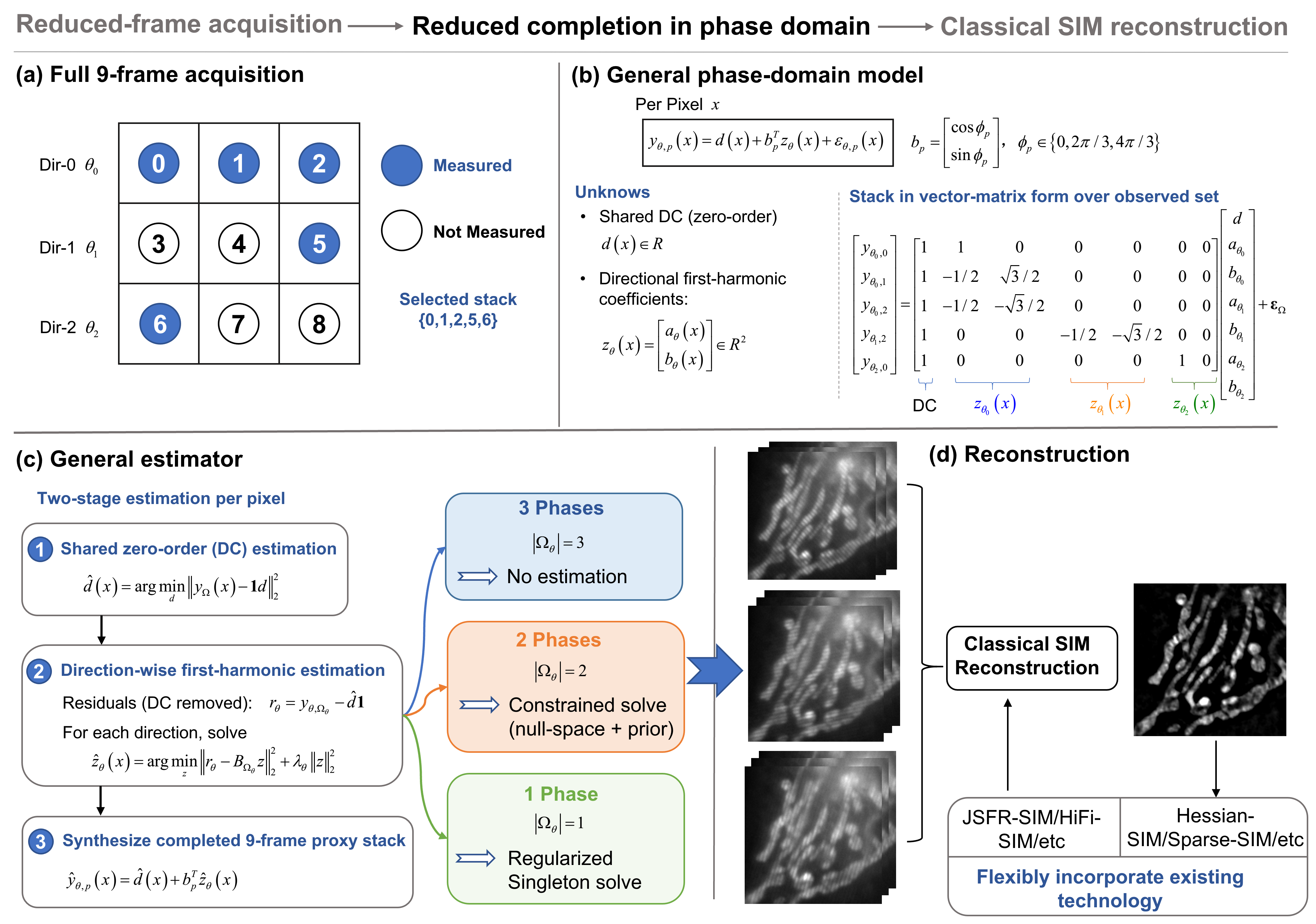}
\caption{Workflow of model-consistent reduced-frame SIM completion. In validation experiments, a fixed subset is retained from a fully sampled raw stack; in reduced-acquisition experiments, the microscope acquires only the selected frames. URA-SIM converts the available frames into a pipeline-compatible completed stack by estimating a shared zero-order field and applying modality-specific phase completion. The completed stack is then reconstructed with the corresponding SIM pipeline.}
\label{fig:workflow}
\end{figure}

For 3-direction, 3-phase linear SIM, the completion model follows the standard sinusoidal illumination form
\begin{equation}
 y_{\theta,p}(x) = \left[h * \left(o(x)\, s_{\theta,p}(x)\right)\right](x) + \varepsilon_{\theta,p}(x),
 \label{eq:forward_spatial}
\end{equation}
with
\begin{equation}
 s_{\theta,p}(x) = 1 + m_{\theta}(x) \cos\!\left(2\pi k_{\theta}\!\cdot\!x + \psi_{\theta}(x) + \phi_p\right).
 \label{eq:illumination}
\end{equation}
Collecting terms with respect to the phase step $\phi_p$ gives
\begin{equation}
 y_{\theta,p}(x) = d(x) + u_{\theta}(x)\cos\phi_p + v_{\theta}(x)\sin\phi_p + \varepsilon_{\theta,p}(x).
 \label{eq:phase_model_directional_dc}
\end{equation}
The derivation is given in Appendix~A. In the online algorithm, $d(x)$ is replaced by a reduced-data estimate $\hat d(x)$ and the residual phase dependence is represented as
\begin{equation}
 y_{\theta,p}(x) = \hat d(x) + b_p^{\top}z_{\theta}(x) + \eta_{\theta,p}(x),
 \qquad
 b_p=\begin{bmatrix}\cos\phi_p\\ \sin\phi_p\end{bmatrix},
 \quad
 z_{\theta}(x)=\begin{bmatrix}u_{\theta}(x)\\ v_{\theta}(x)\end{bmatrix}.
 \label{eq:phase_model_compact}
\end{equation}
For the three-phase protocol used in the 2D-SIM experiments, $\phi_p\in\{0,2\pi/3,4\pi/3\}$. In the current implementation, a practical shared zero-order estimate is computed from the available reduced-frame set $S$ as
\begin{equation}
 \hat d(x) = \alpha \cdot \frac{1}{|S|}\sum_{i\in S} y_i(x),
 \label{eq:dc_estimator}
\end{equation}
where $\alpha$ is a fixed shrinkage factor. This fixed-DC step is the main approximation of the 2D completion model; the exact zero-order field is not jointly optimized with the directional harmonics.

\subsection{2D-SIM reduced-frame completion}
For 3-direction, 3-phase 2D-SIM, the reduced-frame problem is to infer missing phase samples within each illumination direction so that the completed stack matches the 9-frame layout expected by the classical reconstruction pipeline. Let $\Omega_{\theta}\subseteq\{0,1,2\}$ denote the observed phase indices for direction $\theta$. After subtracting the practical shared zero-order estimate $\hat d(x)$, the residual phase dependence is modeled in the two-dimensional first-harmonic basis $\{\cos\phi,\sin\phi\}$.

The baseline estimator is a per-pixel, per-direction ridge fit, with no iterative object-level inverse problem at runtime. The residual definition, objective function, and closed-form solution are given in Appendix~C.

The three phase-availability cases are handled separately. A fully observed direction is copied from the raw data and is not refit. A one-phase direction is underdetermined and is completed by the minimum-energy ridge branch; the implementation parameterizes this branch as a shrinkage gain on the one-phase residual, equivalent to a ridge penalty as shown in Appendix~D. A two-phase direction is identifiable in the fixed-DC two-coefficient ridge model, but the high-fidelity implementation uses a stabilized branch to reduce sensitivity to the shared zero-order estimate. This branch augments the local fit with a local DC term, applies a null-space penalty to the augmented system, and adds an analytic prior calibrated from the observed carrier structure.

Once $\hat z_{\theta}$ has been estimated, the full three-phase directional triplet is synthesized as
\begin{equation}
 \hat y_{\theta,q}(x)=\hat d(x)+b_q^{\top}\hat z_{\theta}(x),
 \qquad q\in\{0,1,2\}.
 \label{eq:surrogate_reconstruction}
\end{equation}
Applying this rule to all three directions yields a completed nine-frame stack. In the 2D incomplete-direction workflow, the fitted triplet may rewrite an observed singleton frame as part of a coherent directional reconstruction; by contrast, the 3D-SIM and NL-SIM completion procedures described below explicitly restore observed frames after synthesis.

\subsection{3D-SIM reduced-frame completion}
For standard 3D-SIM, the full acquisition contains three illumination angles and five phase steps, giving 15 raw frames. The reduced-frame completion follows the same phase-domain principle as the 2D case, but uses the phase basis required by five-phase 3D-SIM processing. For each illumination angle, the available phase volumes are fit in the five-component basis
\begin{equation}
 \{1,\cos\phi,\sin\phi,\cos 2\phi,\sin 2\phi\},
 \label{eq:five_component_phase_basis}
\end{equation}
which contains the DC term, first harmonic, and second harmonic. Missing phase volumes are synthesized from the fitted phase-harmonic representation to form a 15-frame stack, and the originally observed phase volumes are restored in the completed stack before reconstruction. 

\subsection{Nonlinear SIM reduced-frame completion}
The nonlinear SIM experiment uses a 25-frame acquisition format with five illumination directions and five phase steps. Because nonlinear SIM requires higher harmonic content than linear three-phase 2D-SIM, the completion model again uses the five-component phase basis in \eqref{eq:five_component_phase_basis}. For each illumination direction, a direction-wise ridge fit estimates the phase-harmonic coefficients from the available phase samples, synthesizes the missing phases, and restores the observed raw frames in the completed 25-frame stack.

\subsection{Integration with reconstruction pipelines}
\subsubsection{Classical 2D-SIM pipeline}
The 2D completed stack is reconstructed with the joint-space-and-frequency reconstruction approach used in JSFR-SIM \cite{wang2022jsfrsim,wang2023rapid}, which expects a 3-direction, 3-phase input. The compatibility claim is limited but important: the completion stack is constructed on the same DC-plus-first-harmonic phase manifold used by classical three-phase SIM demodulation. URA-SIM therefore does not replace order separation; it supplies a pipeline-compatible phase-completed raw stack.

\subsubsection{Open-3DSIM pipeline}
The completed 15-frame 3D-SIM stack is reconstructed with Open-3DSIM \cite{cao2023open}. URA-SIM therefore acts only before the reconstruction pipeline: it supplies a phase-completed raw stack with the expected angle-phase layout, while Open-3DSIM performs the conventional 3D-SIM reconstruction.

\subsubsection{Nonlinear SIM pipeline}
The completed 25-frame nonlinear SIM stack is passed to the JSFR-NL-SIM reconstruction pipeline \cite{zhang2025highspeed}. As in the 3D-SIM case, the role of URA-SIM is limited to constructing a pipeline-compatible completed raw stack; the subsequent nonlinear SIM reconstruction is performed by the established pipeline.

\section{Results}
\subsection{Reduced-frame 2D-SIM reconstruction on calibration and biological samples}
We first evaluated whether the reduced-frame completion model preserves resolvable structure across both calibration and biological specimens. The experiment uses a R5--R9 progression from fully sampled nine-frame acquisitions, where R5 denotes reconstruction from five retained raw frames and R9 denotes the matched full nine-frame reconstruction from the same acquisition. Because providing all nine raw frames reduces the workflow to the conventional full-frame reconstruction, the R9 output is treated as the full-frame ground-truth reference for the reduced-count comparison.

\begin{figure}[p]
\centering
\includegraphics[width=\textwidth]{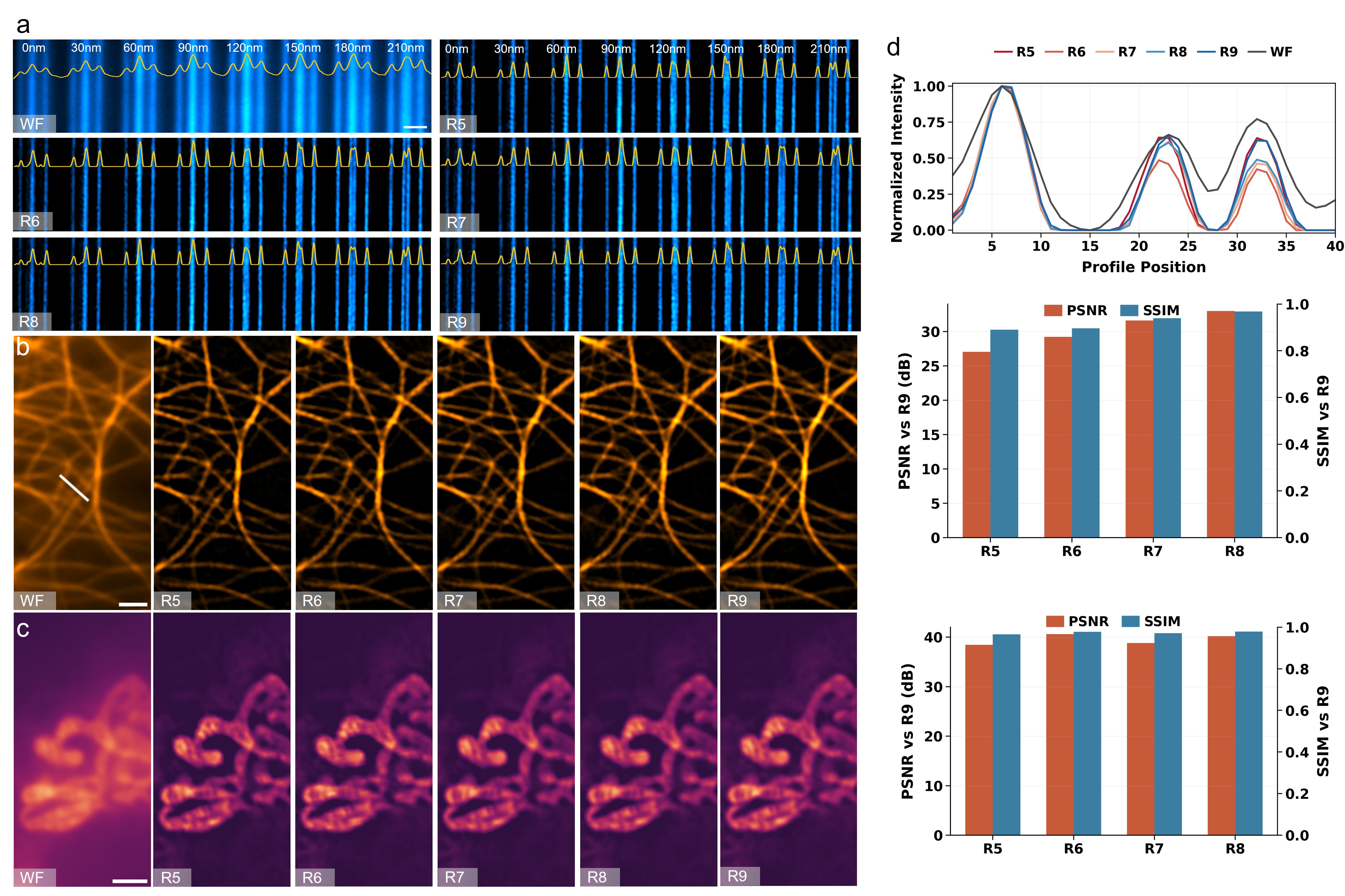}
\caption{\textbf{Reduced-frame URA-SIM reliably reconstructs calibration and biological 2D-SIM samples.}
\textbf{a,} Argolight resolution target reconstructed from wide-field (WF) data and from R5--R9 URA reconstruction. Line profiles are shown above each image region to compare the recovery of the periodic target structure across nominal lateral spacings from 0 to 210\,nm.
\textbf{b,} Microtubule ROI reconstructed as WF and R5--R9 with URA reconstruction. The white line in the WF panel indicates the position used for the profile analysis in \textbf{d}.
\textbf{c,} Mitochondria ROI reconstructed as WF and R5--R9 with URA reconstruction.
\textbf{d,} Quantitative comparison for the biological ROIs. The upper panel shows the normalized intensity profile along the line indicated in \textbf{b}, comparing WF and R5--R9 with URA reconstruction. The middle and lower panels report PSNR and SSIM relative to the matched R9 reconstruction for the microtubule and mitochondria ROIs, respectively.}
\label{fig:argolight_bio_r5_r9}
\end{figure}

Figure~\ref{fig:argolight_bio_r5_r9} validates the reduced-frame reconstructions on an Argolight resolution target, a microtubule ROI, and a mitochondria ROI. Using the R9 reconstruction as the ground truth, the Argolight results show that R9 resolves the 120-nm line pairs, and the R5--R8 reconstructions preserve the same resolvable features with consistent peak registration in the intensity profiles. These observations demonstrate the resolution reliability of the reduced-frame reconstruction algorithm, while the absence of noticeable reconstruction artifacts further supports its image fidelity.
In the microtubule ROI, the full-frame-relative PSNR increases from 27.05\,dB at R5 to 29.22\,dB at R6, 31.61\,dB at R7, and 33.00\,dB at R8, while SSIM increases from 0.890 to 0.896, 0.939, and 0.968, respectively. In the mitochondria ROI, the corresponding PSNR values remain high across the reduced-frame series (38.43, 40.62, 38.79, and 40.20\,dB for R5--R8), and SSIM increases from 0.965 at R5 to 0.977 at R6, 0.972 at R7, and 0.979 at R8. Together with the image panels, these ROI metrics show that the reduced-frame outputs suppress the broad wide-field background while retaining filament continuity and local branching structure, and that the reconstructions closely approach the matched R9 reference in both pixelwise accuracy and structural similarity.

\subsection{COS7 mitochondria reduced-frame method comparison}
Having established the basic R5--R9 behavior on calibration and biological samples, we next compared reduced-frame reconstruction methods on live-cell COS7 mitochondria labelled with Pkmito Orange. This experiment tests whether URA-SIM remains competitive on realistic 2D-SIM cellular data rather than only on calibration targets. The cells were imaged with 561\,nm excitation, 30\,ms exposure for each raw image, and 30\,mW laser power. From each nine-frame SIM acquisition, we retained matched subsets containing five, six, seven, or eight raw frames as the reduced-frame inputs. These R5--R8 inputs were reconstructed with three methods: the proposed URA reconstruction, a spatial-domain Bayesian reduced-frame reconstruction, and a frequency-domain reduced-frame reconstruction. The conventional SIM reconstruction from all nine raw frames was used as the matched super-resolution reference.

\begin{figure}[p]
\centering
\includegraphics[width=\textwidth]{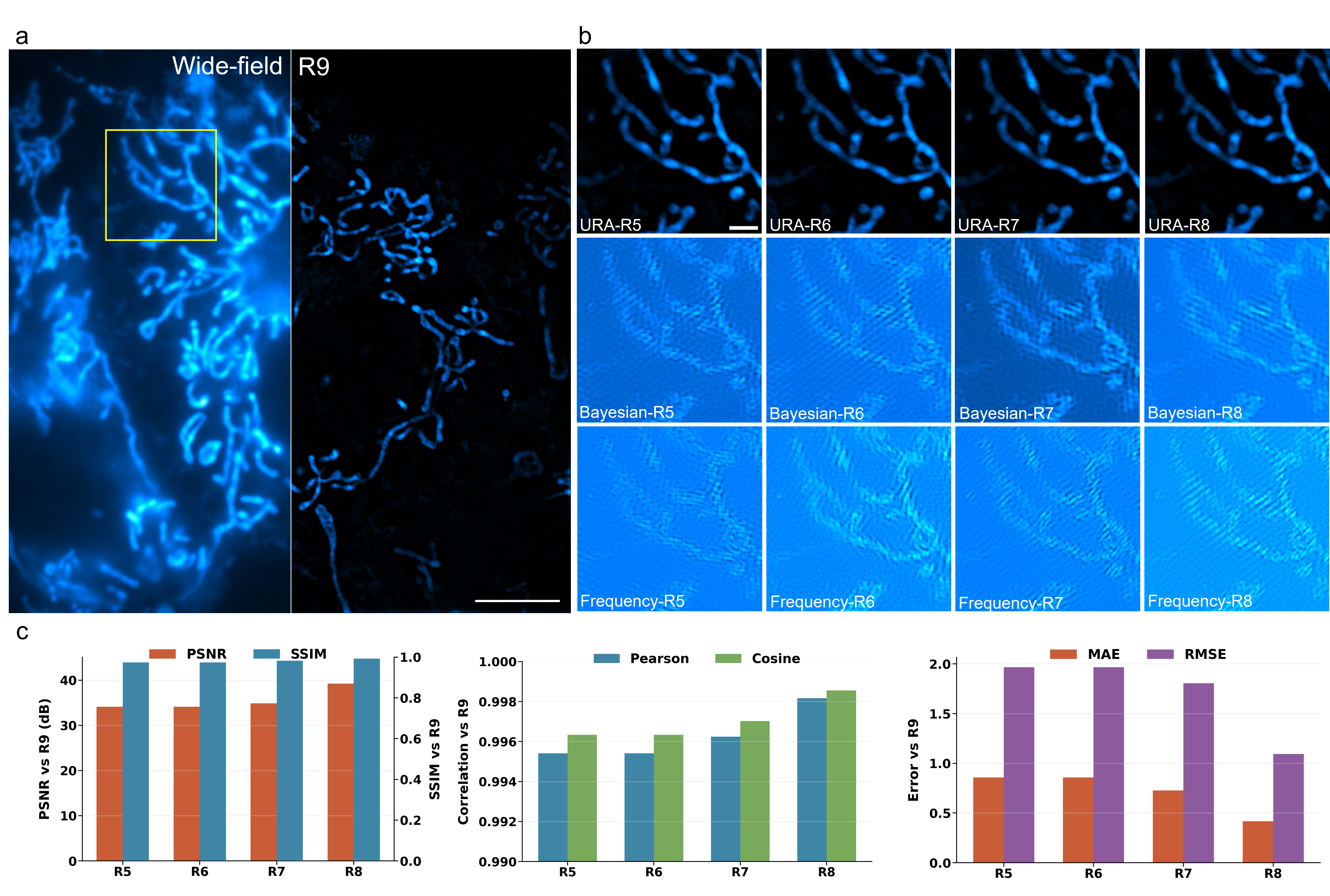}
\caption{\textbf{COS7 mitochondria reduced-frame reconstruction comparison.}
\textbf{a,} Full-field COS7 mitochondria image acquired with Pkmito Orange labelling, 561\,nm excitation, 30\,ms exposure, and 30\,mW laser power. The left half shows the wide-field image, and the right half shows the classical SIM reconstruction from all nine raw frames, used as the matched R9 ground-truth reference. The yellow and white boxes mark the local ROIs used for visual comparison.
\textbf{b,} Local super-resolution reconstructions from reduced-frame inputs containing five to eight raw frames. The first row shows URA reconstructions for R5--R8, the second row shows spatial-domain Bayesian reduced-frame reconstructions, and the third row shows frequency-domain reduced-frame reconstructions.
\textbf{c,} Quantitative comparison of the URA R5--R8 reconstructions against the matched R9 reference, reporting PSNR/SSIM, Pearson/cosine similarity, and MAE/RMSE.}
\label{fig:cos7_mito_compare}
\end{figure}

Figure~\ref{fig:cos7_mito_compare}(a) places the comparison in the full cellular field. The left half shows the wide-field image, while the right half shows the classical SIM reconstruction used as the R9 ground truth, with the yellow boxed regions defining the local comparison ROIs. Figure~\ref{fig:cos7_mito_compare}(b) then compares the local super-resolved outputs for URA, Bayesian reconstruction, and frequency-domain reconstruction across R5--R8. The URA sequence preserves the main mitochondrial tubules with the strongest visual agreement to the R9 reference, whereas the Bayesian and frequency-domain outputs retain more residual stripe or ripple-like texture in the crop. Quantitatively, the URA outputs in Fig.~\ref{fig:cos7_mito_compare}(c) improve as additional raw frames are included: PSNR increases from 34.12\,dB at R5/R6 to 34.87\,dB at R7 and 39.22\,dB at R8, while SSIM increases from 0.9743 at R5/R6 to 0.9818 at R7 and 0.9924 at R8. The same trend is observed in correlation and error metrics, with Pearson/cosine similarity increasing from 0.9954/0.9963 at R5/R6 to 0.9982/0.9986 at R8, and MAE/RMSE decreasing from 0.857/1.967 to 0.417/1.094.

\subsection{Live-cell COS7 mitochondria imaging across acquisition regimes}
\begin{figure}[p]
\centering
\includegraphics[width=\textwidth,height=0.86\textheight,keepaspectratio]{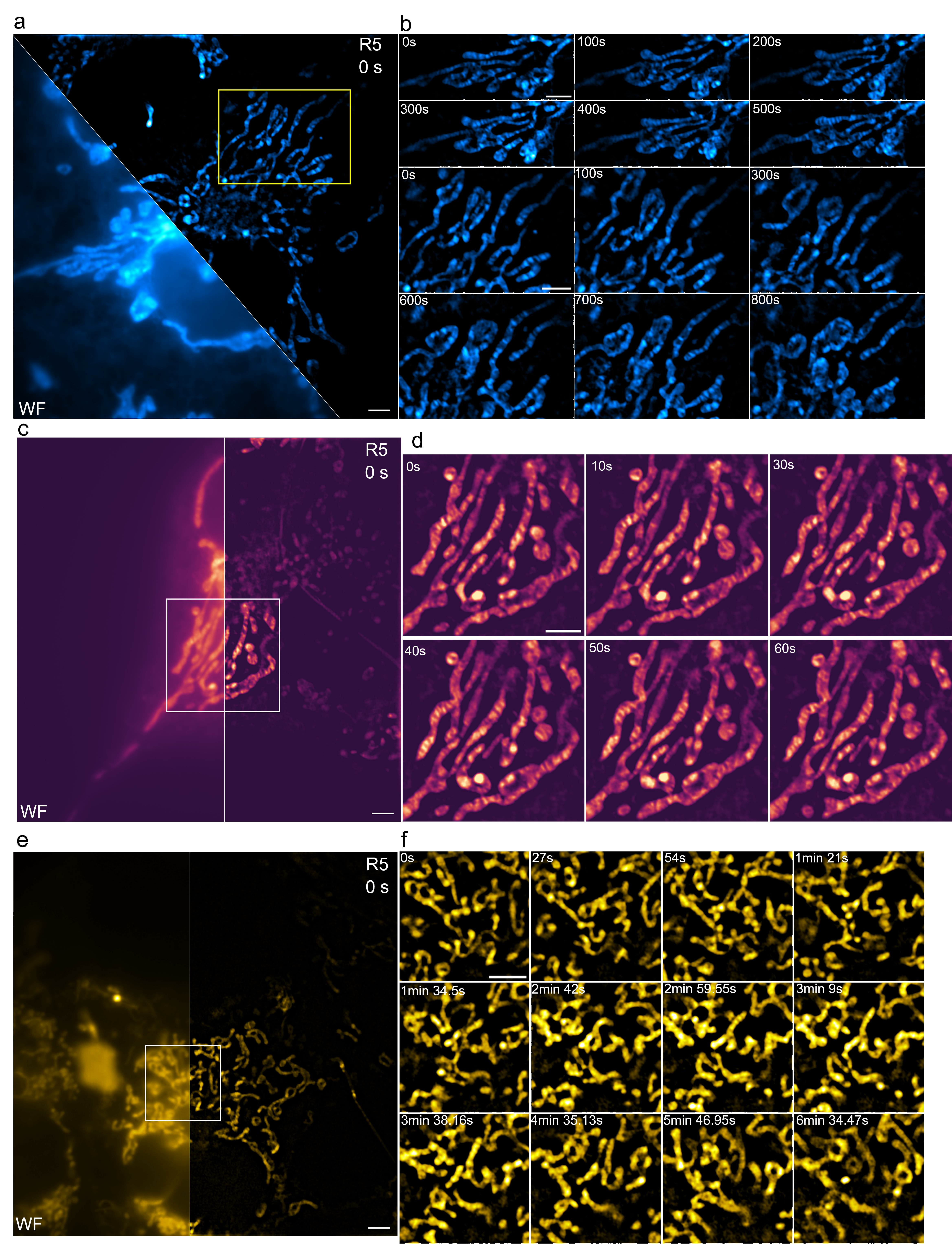}
\caption{\textbf{Live-cell COS7 mitochondria imaging with fixed R5 reconstruction under different temporal sampling regimes.}
\textbf{a,c,e,} Live-cell COS7 mitochondria acquisitions labelled with Pkmito Orange and imaged with 561\,nm excitation, 30\,ms exposure, and 30\,mW laser power. \textbf{a,} Time-lapse acquisition with one super-resolved frame every 20\,s for 40 frames. \textbf{c,} Short-interval acquisition with one super-resolved frame every 1\,s for 60 frames. \textbf{e,} Long continuous acquisition with 1461 super-resolved frames. \textbf{b,d,f,} Corresponding magnified ROIs from \textbf{a,c,e}, respectively, showing mitochondrial dynamics over the indicated time points.}
\label{fig:mitoliveexp}
\end{figure}

We then moved from retrospective subset validation to actual fixed-R5 live-cell imaging. The purpose of this experiment was to test whether URA-SIM can support dynamic mitochondrial observation when only five raw frames are acquired for each reconstructed time point. Unlike the retrospective R5--R8 validation against matched nine-frame references, these live-cell experiments used a fixed reduced-acquisition workflow. All three experiments used COS7 cells labelled with Pkmito Orange, 561\,nm excitation, 30\,ms exposure, and 30\,mW laser power. Figure~\ref{fig:mitoliveexp}(a,b) shows a time-lapse series acquired every 20\,s, yielding 40 super-resolved frames. Figure~\ref{fig:mitoliveexp}(c,d) shows a short-interval acquisition acquired every 1\,s for 60 super-resolved frames. Figure~\ref{fig:mitoliveexp}(e,f) shows a longer continuous acquisition containing 1461 super-resolved frames. In each case, the wide-field view identifies the cellular field and ROI position, while the corresponding magnified panels show that local mitochondrial tubules remain resolvable over the sampled time course.

\subsection{Transfer to 3D-SIM}
After validating the 2D-SIM workflow on our calibration and live-cell experiments, we tested whether the same reduced-frame completion principle can be extended to standard 3D-SIM. The examples use raw datasets from the Open-3DSIM study \cite{cao2023open}: the OMX\_Argolight calibration sample and the OMX\_COS7\_Nup sample. In this setting the full acquisition contains three illumination angles and five phase steps, giving 15 raw frames. For validation, URA-SIM completes reduced 11--14 frame inputs retained from the full stacks into 15-frame-compatible stacks and then reconstructs them with the Open-3DSIM pipeline. The conventional 3D-SIM reconstruction from all 15 raw frames is used as the matched reference for evaluating the reduced-frame reconstructions.

\begin{figure}[p]
\centering
\includegraphics[width=\textwidth]{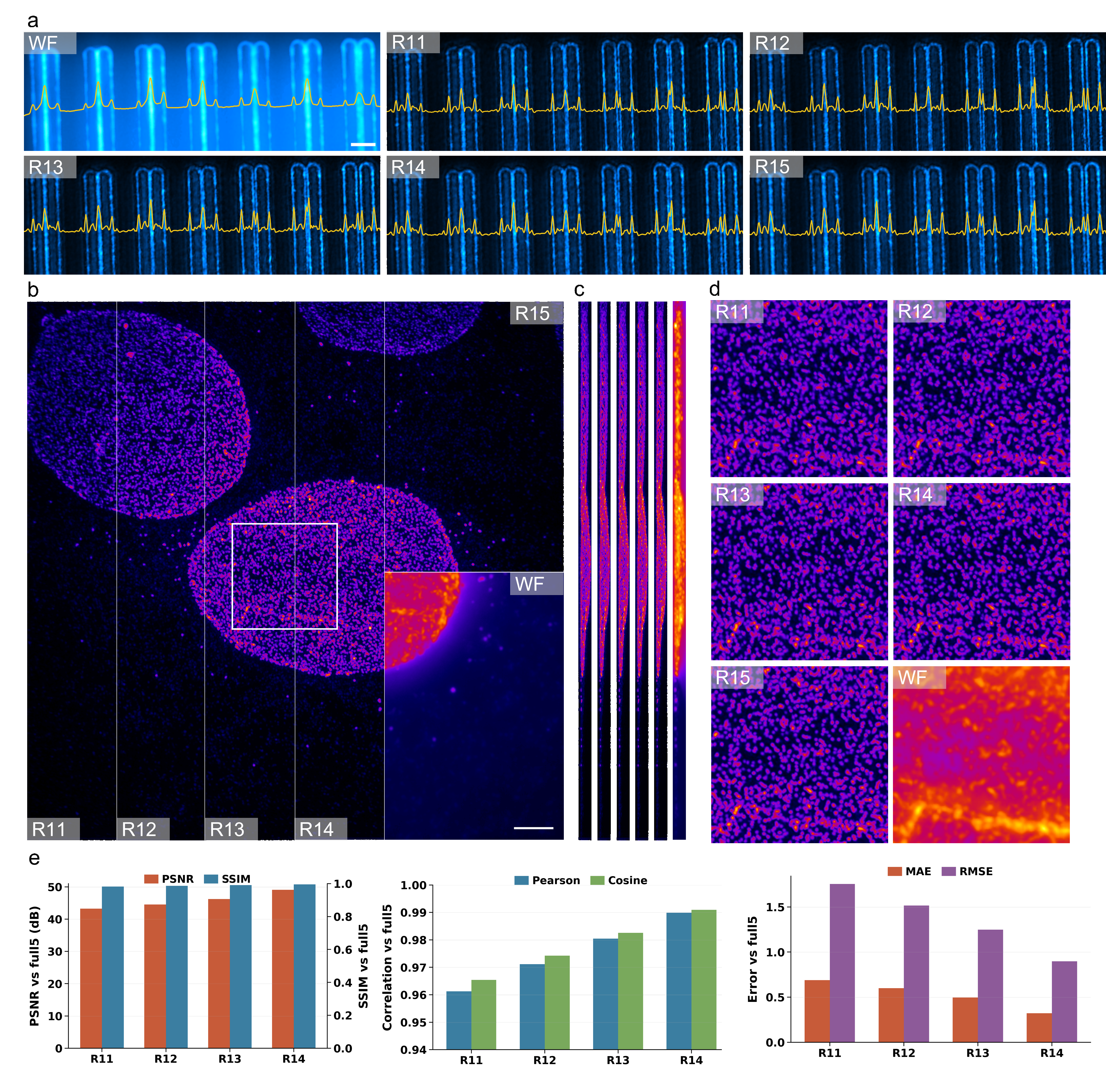}
\caption{\textbf{URA-SIM reduced-frame reconstruction for public Open-3DSIM 3D-SIM data.}
\textbf{a,b,} The samples are from the Open-3DSIM study \cite{cao2023open}. \textbf{a,} OMX\_Argolight standard sample and \textbf{b,} OMX\_COS7\_Nup sample, shown as x-y maximum-intensity projections of SR reconstructions obtained from 11--15 raw frames with URA-SIM completion and Open-3DSIM reconstruction. The reduced-frame reconstructions show comparable resolution to the 15-frame reference without obvious reconstruction artifacts.
\textbf{c,} x-z maximum-intensity projection of the OMX\_COS7\_Nup sample, showing that 11--14 frame reduced reconstruction reaches axial resolution comparable to the 15-frame reconstruction without obvious artifacts.
\textbf{d,} Magnified local ROI from the white boxed region in \textbf{b}.
\textbf{e,} Quantitative comparison of R11--R14 reconstructions against the matched 15-frame 3D-SIM reconstruction. PSNR/SSIM improve from 43.25\,dB/0.9826 at R11 to 44.53\,dB/0.9866 at R12, 46.20\,dB/0.9905 at R13, and 49.07\,dB/0.9952 at R14. Over the same range, Pearson/cosine similarity increases from 0.9612/0.9654 to 0.9899/0.9909, while MAE/RMSE decreases from 0.690/1.754 to 0.322/0.897.}
\label{fig:3dura}
\end{figure}

Figure~\ref{fig:3dura} shows that the reduced-frame 3D-SIM reconstructions preserve both lateral and axial structure. For OMX\_Argolight and OMX\_COS7\_Nup, the x-y maximum-intensity projections from 11--15 frames show comparable apparent resolution and no obvious reconstruction artifacts. The x-z projection of the COS7 Nup sample likewise shows that the 11--14 frame reconstructions reach an axial resolution visually consistent with the 15-frame reference. Quantitatively, comparison against the matched 15-frame 3D-SIM reconstruction gives PSNR/SSIM values of 43.25\,dB/0.9826 at R11, 44.53\,dB/0.9866 at R12, 46.20\,dB/0.9905 at R13, and 49.07\,dB/0.9952 at R14. Pearson/cosine similarity also increases from 0.9612/0.9654 at R11 to 0.9899/0.9909 at R14, while MAE/RMSE decreases from 0.690/1.754 to 0.322/0.897.

\subsection{Transfer to nonlinear SIM}
\begin{figure}[p]
\centering
\includegraphics[width=\textwidth]{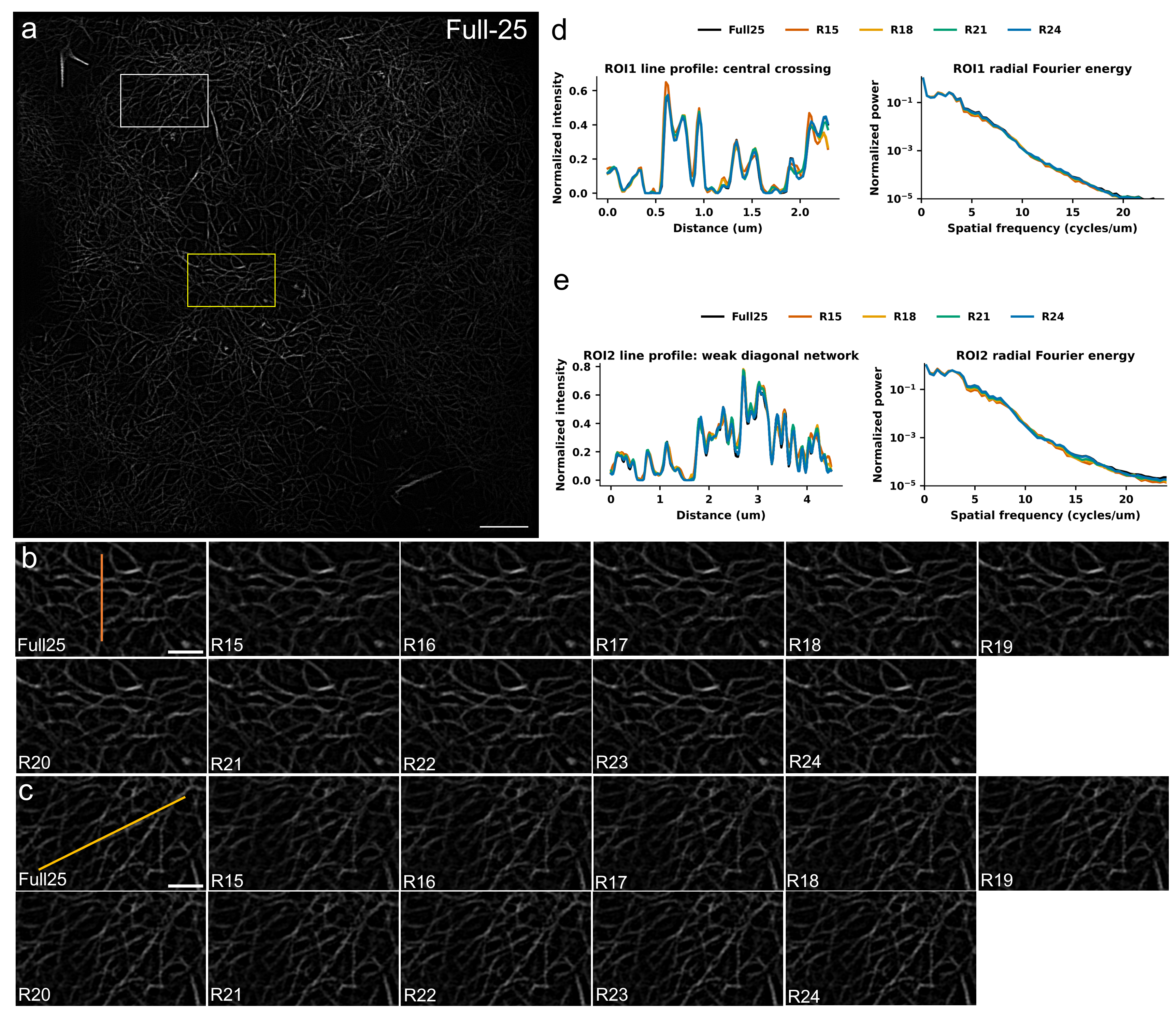}
\caption{\textbf{Reduced-frame reconstruction on public nonlinear SIM data.}
\textbf{a,} Standard 25-frame JSFR-NL-SIM reconstruction with two boxed regions used for the detailed comparisons.
\textbf{b,c,} Cropped reconstructions from the matched 25-frame reference and from reduced-frame reconstructions using 15--24 raw frames. The orange lines indicate the positions used for line-profile analysis.
\textbf{d,e,} Line profiles and radial Fourier-energy curves for ROI1 and ROI2, comparing the 25-frame reference with representative reduced-frame reconstructions using 15, 18, 21 and 24 raw frames. Reduced-frame reconstructions preserve the positions of the dominant filament peaks and progressively recover the high-frequency energy distribution as more phase samples are added.}
\label{fig:nlsim}
\end{figure}

We further asked whether the completion idea can be transferred to public nonlinear SIM data. This experiment is also a transfer test rather than a new acquisition. Nonlinear SIM provides a useful stress test because its raw acquisition contains 25 images arranged as 5 illumination directions and 5 phase steps. More importantly, the phase model is not the same as in linear SIM: the reconstruction must retain the DC term, the first harmonic, and the second harmonic. The implementation used for Fig.~\ref{fig:nlsim} therefore adopts the nonlinear five-component phase basis in \eqref{eq:five_component_phase_basis}, completes missing phase samples direction by direction, and then passes the resulting 25-frame stack to the JSFR-NL-SIM reconstruction pipeline. The standard JSFR-NL-SIM reconstruction from all 25 raw frames is used as the matched reference for evaluating the reduced-frame reconstructions, rather than as an independent biological ground truth.

The nonlinear experiment changes the strength of the claim. In the linear case, the first-harmonic phase model follows directly from the standard sinusoidal SIM forward model. In nonlinear SIM, the five-component basis in \eqref{eq:five_component_phase_basis} is a model-aware extension designed to match the harmonics used by the JSFR-NL-SIM reconstruction pipeline. Figure~\ref{fig:nlsim} shows that this distinction matters in a useful way. The 15-frame reconstruction already recovers the main mitochondrial network, but the crop panels and line profiles reveal residual intensity and contrast deviations. As the raw-frame count increases from 15 to 24, filament locations remain registered to the matched 25-frame JSFR-NL-SIM reconstruction, and the radial Fourier-energy curves converge over the mid- and high-frequency ranges. The corresponding full-field affine SSIM values against the 25-frame reference increase from 0.9054 at 15 frames to 0.9451 at 18 frames, 0.9775 at 21 frames, and 0.9954 at 24 frames; the ROI-gradient affine SSIM follows the same trend, increasing from 0.6572 to 0.7803, 0.9095 and 0.9852, respectively.

These data support a limited but important generalization: phase-domain completion is not restricted to the three-phase linear-SIM basis, provided that the completion model is changed to match the harmonic content expected by the reconstruction engine. The nonlinear result should not be read as evidence that the linear first-harmonic derivation automatically applies to nonlinear SIM. Rather, it shows that the same design principle---complete the raw stack on the correct phase-harmonic manifold, then call a mature reconstruction pipeline---can remain effective when the manifold is expanded to include the second harmonic.

\section{Discussion}

URA-SIM provides a practical route for reducing the raw-frame burden of SIM while retaining compatibility with mature reconstruction pipelines. The central result of this study is that a fixed reduced-frame input can be converted into a pipeline-compatible raw stack by completing the missing phase samples in a low-dimensional phase-harmonic representation. In 2D-SIM, this strategy preserves the resolvable structure of calibration and biological samples across the R5--R9 progression, remains competitive in a COS7 mitochondria method comparison, and supports fixed-R5 live-cell imaging in which each reconstructed time point is acquired from only five raw frames. These results establish the main use case of the method: stable, high-throughput 2D-SIM reconstruction under a predefined reduced-acquisition protocol.

The effectiveness of URA-SIM follows from the structure of the SIM forward model. For linear 2D-SIM, the phase dependence of each illumination direction is naturally represented by a zero-order term and a sine-cosine first harmonic. The coefficient vector $z_{\theta}(x)$ in \eqref{eq:phase_model_compact} is therefore not an arbitrary latent representation; it is the same phase-harmonic structure used by conventional three-phase demodulation. URA-SIM exploits this structure before reconstruction, estimating a shared zero-order field and completing incomplete directional triplets through branch-specific harmonic fitting.

This design differs from earlier and parallel reduced-image SIM strategies. In saturated patterned-excitation microscopy, the nominal number of required images can be reduced by exploiting redundancy in coupled Fourier-domain equations \cite{heintzmann2003saturated}, while triangle-beam interference SIM reduces the raw-frame burden through a modified two-dimensional illumination geometry \cite{fu2025triangle}. Bayesian reduced-frame reconstruction demonstrated that incomplete phase measurements can be handled probabilistically, but the reconstruction remains sensitive to prior assumptions, illumination-parameter mismatch, and posterior uncertainty when phase diversity is limited \cite{orieux2012bayesian}. Frequency-domain reduced-image reconstruction provides a more direct reduced-frame inverse formulation, but its stability depends on accurate frequency-domain separation and its computational burden can limit runtime use in live-cell workflows \cite{lal2018frequency}. URA-SIM takes a different route: instead of redesigning the illumination geometry or solving a large object-level inverse problem at runtime, it performs local phase-domain completion in a modality-specific harmonic representation and then uses an established SIM reconstruction pipeline for order separation and image formation. This division of labor keeps the online computation lightweight and explicit, while preserving the calibration and optical-transfer-function handling already present in the classical pipeline.

The live-cell experiments are central to the interpretation of the method. Retrospective R5--R8 tests against matched nine-frame references provide controlled evidence that reduced-frame completion preserves image content, but the fixed-R5 acquisitions test the actual intended workflow: fewer raw frames are collected at each time point before reconstruction. The COS7 mitochondria data show that URA-SIM can maintain local mitochondrial morphology over different temporal sampling regimes, including 20\,s interval imaging, 1\,s interval imaging, and a long continuous acquisition. This supports the use of reduced-frame SIM not only as an offline reconstruction exercise, but also as an acquisition strategy for dynamic cellular imaging.

The 3D-SIM and nonlinear SIM experiments show how the same design principle can be transferred when the phase model and reconstruction-pipeline interface are changed appropriately. In the Open-3DSIM experiments, URA-SIM completes reduced 11--14 frame inputs into 15-frame-compatible stacks using a five-component phase basis, and the resulting reconstructions approach the matched 15-frame 3D-SIM reconstruction in both visual quality and quantitative similarity. In the nonlinear SIM experiment, the completion basis is expanded to include the second harmonic required by the JSFR-NL-SIM pipeline. These public-dataset results indicate that URA-SIM is best viewed as a modular completion framework: the phase basis is modality-specific, but the workflow of completing a reduced raw stack before mature reconstruction-pipeline processing is shared.

The present implementation is intentionally tied to fixed reduced-frame protocols, because fixed protocols are the most relevant setting for reproducible live-cell acquisition and stable reconstruction throughput. The shared zero-order estimate and branch-specific completion rules provide a compact implementation that worked well in the experiments reported here. Future implementations can further refine this component by jointly estimating the zero-order and harmonic terms, adding optional hard consistency constraints for observed frames, or incorporating stronger calibration updates when the illumination pattern varies across time. These developments can be added without changing the central architecture of URA-SIM: a transparent phase-domain completion layer coupled to a well-tested SIM reconstruction pipeline.

\section{Conclusion}
This paper presents URA-SIM as a phase-domain completion framework for reduced-frame SIM reconstruction. Under a fixed 3-direction, 3-phase acquisition format, URA-SIM interprets each illumination direction through the first-harmonic basis $\{1,\cos\phi,\sin\phi\}$, estimates a shared zero-order component, and completes incomplete directions through branch-specific harmonic estimation. The completed nine-frame stack is constructed to match the harmonic structure expected by a classical simrecon pipeline.   The 2D-SIM results show that URA-SIM preserves calibration and biological structures across reduced-frame settings, remains competitive on COS7 mitochondria comparison data, and enables fixed-R5 live-cell mitochondrial imaging. Public 3D-SIM and nonlinear SIM datasets further show that the same completion principle can be extended by adapting the phase basis and reconstruction-pipeline interface. Together, these results support URA-SIM as a transparent, model-consistent, and computationally lightweight route from fixed reduced-frame acquisition to mature SIM reconstruction workflows.

\clearpage
\bibliographystyle{unsrt}
\bibliography{ref}

\end{document}